# First-Principles Study of Structural and Electronic Properties of Germanene

Harihar Behera and Gautam Mukhopadhyay

*Department of Physics, Indian Institute of Technology Bombay, Mumbai-400076, India*

**Abstract.** The ground state structural and electronic properties of germanene (the germanium analogue of graphene) are investigated using first-principles calculations. On structure optimization, the graphene-like honeycomb structure of germanene turns out as buckled (buckling parameter $\Delta = 0.635$ Å) in contrast with graphene's planar structure (buckling parameter $\Delta = 0.0$ Å). In spite of this, germanene has similar electronic structure as that of graphene. While corroborating the reported results, we newly predict the in-plane contraction of hexagonal Ge with (thermal) stretching along the "c" axis, akin to a phenomenon observed in graphite.

## INTRODUCTION

Synthesis of a single atomic plane of graphite, i.e., graphene, with covalently bonded honeycomb lattice, has been a breakthrough in the world of materials research [1-4]. Graphene, because of its striking properties, has potentials for many novel applications in nanoelectronic devices [1-4]. However, "graphenium" microprocessors are unlikely to appear in near future, since replacement of silicon electronics remains a hurdle [4]. In order to be compatible with Si based microtechnology, Si and Ge based nanotechnology is highly promising. Recently, the electronic properties of two-dimensional hexagonal crystals of Si and Ge, so called silicene and germanene, respectively, have been theoretically studied [5-8] and there are reports of strips of silicene having been successfully epitaxially grown on Ag(110) and Ag(100) surfaces [9-11]. However, germanene remains a hypothetical material.

Recently, we have reported [12] our first-principles study on graphene and silicene using the highly accurate density functional theory based full-potential (linearized) augmented plane wave plus local orbital (FP(L)APW+lo) method [13,14], which is a descendant of FP-LAPW method [15]. Here, we report our calculated results on the ground state properties of germanene based on the same method [13,14] and compared our results with some recent results from other methods [5-8] such as the pseudo-potential and projector augmented-potential (PAW-pot) methods.

## CALCULATION METHODS

The calculations have been performed by employing the FP-(L)APW+lo method [13,14] as implemented in the elk-code [16]. For exchange-correlation (xc-) functional, we used the local density approximation (LDA) [17]. The plane wave cutoff of $|\mathbf{G+k}|_{max} = 8.0/R_{mt}$ ($R_{mt}$ is the smallest muffin-tin radius in the unit cell) is used for the plane wave expansion of the wave function in the interstitial region. For structural calculations the k point mesh of (20×20×1) and for band structure and density of states calculations k point mesh of (30×30×1) are used for the convergence of relevant quantities. The convergence of total energy was chosen as less than 3 μeV/unit cell between two steps and the ionic relaxation was carried out until the forces acting on the atoms decreased below 2.5 meV/Å. To simulate the two-dimensional hexagonal structure of germanene, three-dimensional hexagonal supercell with large values of the "c" ( c = 30, 40 a.u.) parameter were used.

## RESULTS AND DISCUSSIONS

The calculated ground state results for germanene along with other relevant information are given in Table 1. As expected, our computed value of the in-plane lattice constant "a" is larger than those of our results for graphene (a = 2.4449 Å) and silicene (a=

3.8081 Å) for the same value of c= 40 a.u. [12]. Analogous to our previous prediction [12] for contraction of "a" value with increasing "c" value for hexagonal C and hexagonal Si, here we predict a similar effect for germanene as shown in Table 1. For "c" → ∞, the extrapolated value of "a" turns out to be a = 3.9319 Å. Given this contraction of "a" arising out of the increasing value of the "c" parameter used, and the basic differences between different other methods, the deviations of our results with others are reasonable.

The band structure and total density of states of germanene in the buckled structure with a buckling parameter $\Delta$ = 0.635Å, which we obtained from the ionic relaxation within LDA with c = 40 a.u., are shown in Fig. 1. In the buckled structure, the locations of the alternating atoms of the hexagonal lattice are in two different parallel planes, i.e., basal planes in this case; the buckling parameter $\Delta$ is the perpendicular distance between these two planes. As seen in Fig. 1, the energy bands of germanene strongly resembles with those of graphene [6, 12]. In particular, the energy gap is zero at the K-point of the Brillouin zone and the one-particle energy dispersion around this point is linear. This is the property of the so-called Dirac-cone [2-4] in graphene that is mostly responsible for its unusual exotic properties. The present results for germanene corroborate the reported results [5-8] based on other methods.

**Table 1.** Calculated ground state results of germanene. The in-plane lattice constant $|\mathbf{a}_1| = |\mathbf{a}_2|$ = a, the buckling parameter $\Delta$, and energy gap $E_G$ at K-point are compared with reported values. The value of the out-of-plane lattice parameter $|\mathbf{a}_3|$ = c used for calculation of "a" are given under the table-head "c" wherever they are available.

|  | xc-functional | a (Å) | $\Delta$ (Å) | $E_G$ (eV) | c (a.u) | Remark |
|---|---|---|---|---|---|---|
| Germanene | LDA | 3.9455 | 0.635 (used) | 0 | 30 | This work |
|  | LDA | 3.9421 | 0.635 | 0 | 40 | This work |
|  | LDA | 3.97 | 0.64 | 0 |  | PAW-pot [6] |
|  | LDA-HGH | 3.9204 | 0.6220 | 0 | 80 | Ps-pot [7] |
|  | LDA | 4.034 | 0.0 (used) | 0 |  | PAW-pot [5] |
|  | LDA | 4.02 | 0.0 (used) | 0 |  | Ps-pot [8] |

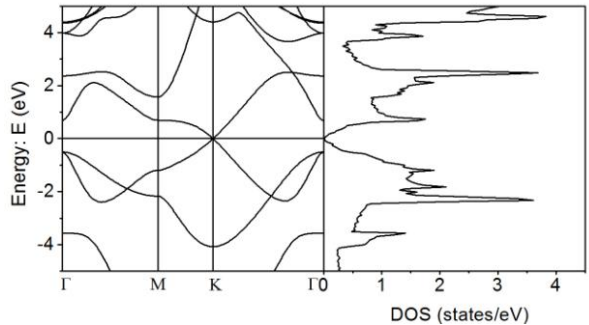

**FIGURE 1** The band structure and total density of states (DOS) of buckled germanene. The Fermi level is set at 0 eV.

## CONCLUSIONS

In this first-principles study of germanene employing the FP-(L)APW+lo method, we found that the buckled germanene is more stable than the planar one and yet has an electronic structure resembling that of graphene with a linear energy dispersion around the K point. We also newly predict the in-plane contraction of hexagonal Ge with (thermal) stretching along the "c"-axis, akin to a phenomenon observed in graphite.